\documentclass[journal,twocolumn]{IEEEtran}
\usepackage{epsfig,makeidx,color}
\usepackage{amsmath,amssymb,bbm}
\usepackage{cite,graphicx,lipsum}
\usepackage{enumerate}
\usepackage[switch,pagewise]{lineno}
\usepackage{hyperref}
\hypersetup{
        colorlinks = true,
        citecolor=blue,
}
\def\cQ{{\cal Q}}

\def\rT{{\rm T}}

\def\uE{{\mathbb E}}

\newtheorem{mylemma}{\bf Lemma} 

\def\be{ \begin{equation} }
\def\ee{ \end{equation} }
\def\bea{ \begin{eqnarray} }
\def\eea{ \end{eqnarray} }

\def\bC{{\bf C}}

\def\bI{{\bf I}}

\def\b0{{\bf 0}}

\def\cQ{{\cal Q}}

\ifCLASSOPTIONonecolumn
  \newcommand{\figwidth}{0.50\columnwidth}
  
\else
  \newcommand{\figwidth}{0.85\columnwidth}
  
\fi

\begin{document}

\title{Sliding Network Coding for URLLC}

\author{Jinho Choi\\
\thanks{The author is with
the School of Information Technology,
Deakin University, Geelong, VIC 3220, Australia
(e-mail: jinho.choi@deakin.edu.au).
This research was supported
by the Australian Government through the Australian Research
Council's Discovery Projects funding scheme (DP200100391).}}


\maketitle
\begin{abstract}
In this paper, we propose a network coding (NC) based approach  
to ultra-reliable low-latency communication
(URLLC) over erasure channels.
In transmitting multiple data packets,
we demonstrate that the use of random NC can improve
the reliability in terms of decoding error probability,
while it incurs a longer decoding delay than a well-known
$K$-repetition. To avoid a long decoding delay,
we consider a sliding NC (SNC) design that allows a highly
reliable transmission of 
each data packet with a guaranteed decoding delay.
A few design examples are derived and their decoding error
rates are analyzed. Through the analysis,
we can show that the decoding error rate of SNC
is much lower 
than that of $K$-repetition at the same spectral efficiency,
which means a more reliable transmission can be achieved
using SNC than $K$-repetition.
\end{abstract}

\begin{IEEEkeywords}
URLLC; Network coding; Sliding window
\end{IEEEkeywords}

\ifCLASSOPTIONonecolumn
\baselineskip 28pt
\fi

\section{Introduction}
In
the fifth generation (5G)
technology standard for celluar systems \cite{Shafi17} \cite{Agiwal_5g},
a number of new services and applications are to be supported. 
Among them, there are mission-critical applications
(including  industrial automation and autonomous vehicles)
that are to be supported by ultra-reliable low-latency
communication (URLLC) \cite{Sutton19} \cite{TR38824_URLLC}\cite{5GAmericasURLLC}. 
In general, 
URLLC is considered
to be the most challenging task in 5G and future
wireless  networks 
due to
two  ambitious  requirements,
namely, high-reliability and low-latency, to be satisfied simultaneously
for short-packet transmissions 
\cite{Popovski_urllc} \cite{Bennis_URLLC} \cite{pokhrel2020towards}.
According to \cite{3GPP_TS22.146},
5G has identified several scenarios of
URLLC with performance
requirements. 
For example, for factory automation, the actuation of industrial devices has stringent performance requirements 
such as a latency of 1 millisecond (ms) and reliability of 99.9999\%.

In general, in order to achieve
a high reliability, hybrid automatic repeat request
(HARQ) protocols can be used \cite{WickerBook} \cite{LinBook}.
In HARQ as a link-layer protocol for a peer-to-peer communication,
coded packets are transmitted
by a transmitter, and some of them are
re-transmitted 
if a receiver is unable to decode them due to channel
fading, interference, or any other reasons. 
To enable re-transmissions, the feedback from the
receiver is sent to the transmitter.
In general, HARQ protocols can achieve a high reliability. 
However, if there are frequent re-transmissions,
the decoding delay of packets can be long as the transmitter
needs to wait until it receives a feedback signal.
There are variants to reduce decoding delay \cite{Makki19}
\cite{Strodthoff19} based on decoding error prediction.

An effective means to lower the delay in HARQ is to 
exploit transmit diversity, e.g.,
the same packet can be transmitted a number of times,
called $K$-repetition \cite{3GPP_HARQ1} \cite{Pocovi18}.
Since the probability of successful transmission increases
with the number of repetitions, $K$, 
the number of re-transmissions can decrease, which leads
to a short decoding delay at the cost of the spectral
efficiency by a factor of $K$. This is often acceptable 
to meet a stringent delay constraint when
the bandwidth
is plentiful \cite{Karzand17}.

The notion of 
network coding (NC) has been introduced for
efficient routing of multicast traffic \cite{Ahlswede}
\cite{Ho06}
and extended to various applications \cite{Fragouli06} \cite{Chou07}
\cite{Sund11}.
Among them,  it is shown in 
\cite{Karzand17} that NC can be used to
meet low delay requirements in 5G.

As in \cite{Karzand17},
in this paper, we propose an approach based on NC
to URLLC (the relationship between the proposed
approach and the approach in \cite{Karzand17} will be explained
in Subsection~\ref{SS:RW}). 
In this approach, NC packets, which are
linear combinations of original
data packets, are transmitted together
with original data packets. 
In particular, NC packets are generated using a
sliding window of
original data packets, and for this reason, the proposed 
approach is called sliding NC (SNC).
This approach allows a receiver to decode a sequence 
of coded packets on-the-fly with a certain specific
decoding delay. As a result, when a transmitter 
generates a sequence of packets at a certain rate
and wishes to deliver
each packet with a guaranteed delay in URLLC applications,
the proposed approach can be used while providing
a high transmission reliability.

The main contributions of the paper
can be summarized as follows:
\emph{i)} the notion of SNC is proposed to transmit
packets in on-the-fly mode 
to meet URLLC requirements
in terms of decoding delay and packet decoding error rate;
\emph{ii)} 
various SNC designs are derived with a delay constraint;
\emph{iii)} the decoding error rate
of SNC designs is analyzed, which
shows that 
the decoding error rate can be significantly low
compared to that of $K$-repetition
at the same level of 
spectral efficiency.

\subsection{Related Works}  \label{SS:RW}

For URLLC, coded short packets are considered as in
\cite{Shirvan} \cite{Durisi16},
where a low decoding error rate is to be achieved
for each short packet transmission.
However, 
if a transmitter has a long message or a sequence of packets
that are generated at a certain rate,
it is necessary to consider streaming codes.
In \cite{Karzand17}, using NC, an approach to generate
a sequence of coded packets, as a streaming
code, is proposed to exploit
the rate-delay trade-off.
In particular, a linear combination of (past) data
packets is inserted after a certain number of original 
data packets,
say $l-1$ packets, where $l \ge 2$,
and transmitted together with original packets.
As a result, the effective code rate becomes
$\frac{l-1}{l}$. In general, 
this approach works well when the channel condition
is not severe. 
If the channel is not reliable for each packet
transmission (as in random access channel
\cite{Singh18} \cite{Choi21}),
the approach in \cite{Karzand17}
cannot provide a sufficiently low decoding error rate. 
In particular, for an erasure channel of an
erasure probability of $\epsilon$, the channel capacity
is $1 - \epsilon$ \cite{CoverBook}. Since
the code rate is $\frac{l-1}{l} = 1 - \frac{1}{l}$
for a positive integer $l$ in \cite{Karzand17},
it is required that $\frac{1}{l} > \epsilon$ for a highly
reliable transmission. If $\epsilon$ is not sufficiently
small due to
hostile channel conditions, 
$l = 2$ (i.e., one NC packet after every one original data
packet) may not ensure a highly reliable transmission
(or a very low decoding error rate). 
Thus, the effective code rate needs to be low 
as that of $K$-repetition \cite{3GPP_HARQ1} \cite{Pocovi18},
which is $\frac{1}{K}$.
The proposed approach in this paper has a low effective code
rate so that a very low decoding error rate can be achieved
by inserting multiple NC packets after every one 
original data packet. In this sense,
the proposed approach can be seen as a generalization of
the approach in \cite{Karzand17}.

\subsection{Organization of the Paper}

The rest of the paper is organized as follows.
In Section~\ref{S:SM}, we present two erasure channel models.
Two different approaches to reliable transmissions over
erasure channels are discussed
in Section~\ref{S:RT}.
In Section~\ref{S:SNC}, we present the proposed approach,
namely SNC, with some design examples.
The decoding error probability of 
the SNC designs introduced in Section~\ref{S:SNC} is analyzed
in Section~\ref{S:PA}.
Simulation results are presented in Section~\ref{S:Sim}
and the paper is concluded in Section~\ref{S:Con}
with a few remarks.

\subsubsection*{Notation}
Matrices and vectors are denoted by upper- and lower-case
boldface letters, respectively. The superscript $\rT$
denotes the transpose. The identity matrix is 
represented by $\bI$.
$\uE[\cdot]$
and ${\rm Var}(\cdot)$
denote the statistical expectation and variance, respectively.
$\cQ(x)$
represents the Q-function, which is defined
as $\cQ(x) = \int_x^\infty \frac{e^{-\frac{z^2}{2}}}{\sqrt{2\pi}} dz$.

\section{Erasure Channel Models}  \label{S:SM}

In this section,
we consider two erasure channel models. 
For convenience, 
assume that time is divided into discrete slots
and a packet can be transmitted within a slot.

\subsection{An Erasure Channel Model for Coded Packets}

Consider a point-to-point channel 
and assume that each packet is a codeword.
From \cite{Polyanskiy10IT} \cite{Durisi16},
the achievable rate of $n$-length code
is given by
\be
R(\rho,n, \epsilon) = \log_2 (1 + \rho)
- \sqrt\frac{ V (\rho)}{n} \cQ^{-1} (\epsilon) +
O\left(\frac{\log_2 n}{n}\right),
        \label{EQ:R_PPV}
\ee
where $\rho$
is the signal-to-noise ratio
(SNR), $\epsilon$ is the (codeword or packet) error probability,
$V(\rho)$ is the channel dispersion that is given by
\be
V(\rho)
= \frac{\rho(2 + \rho)}{(1+ \rho)^2} (\log_2 e)^2 .
\ee
Alternatively,
the  error probability becomes
\be
\epsilon \approx \cQ \left(
\sqrt{\frac{n }{V(\rho)}} 
\left( \log_2 (1+\rho) - \frac{N_{\rm bit}}{n} \right)
\right),
    \label{EQ:ec1}
\ee
where $N_{\rm bit} = Rn$ represents the number of message bits 
per packet
and $n$ can be seen as the number of channel uses.
Here, the code rate, $\frac{N_{\rm bit}}{n}$,
should be lower than the capacity, $\log_2 (1+\rho)$,
for a low error probability.
As a result, the channel 
can be seen as an erasure
channel with the erasure probability $\epsilon$ for 
each packet transmission.

Note that for fading channels,
the right-hand side (RHS) in \eqref{EQ:ec1} is to be
averaged over the SNR, $\rho$, to find the 
average erasure probability
\cite{Yang14}
\cite{Durisi16},
where the SNR is the received SNR that includes
the random channel coefficient.

\subsection{An Erasure Channel Model with Two-Step Random Access} 

For machine-type communication (MTC)
in 5G, 2-step random access has been considered \cite{3GPP_MTC_18}
in order to lower signaling overhead compared to conventional
4-step random access. 
In this subsection, we will show that the resulting channel
of 2-step random access can also be seen as an erasure channel.

\begin{figure}[thb]
\begin{center}
\includegraphics[width=\figwidth]{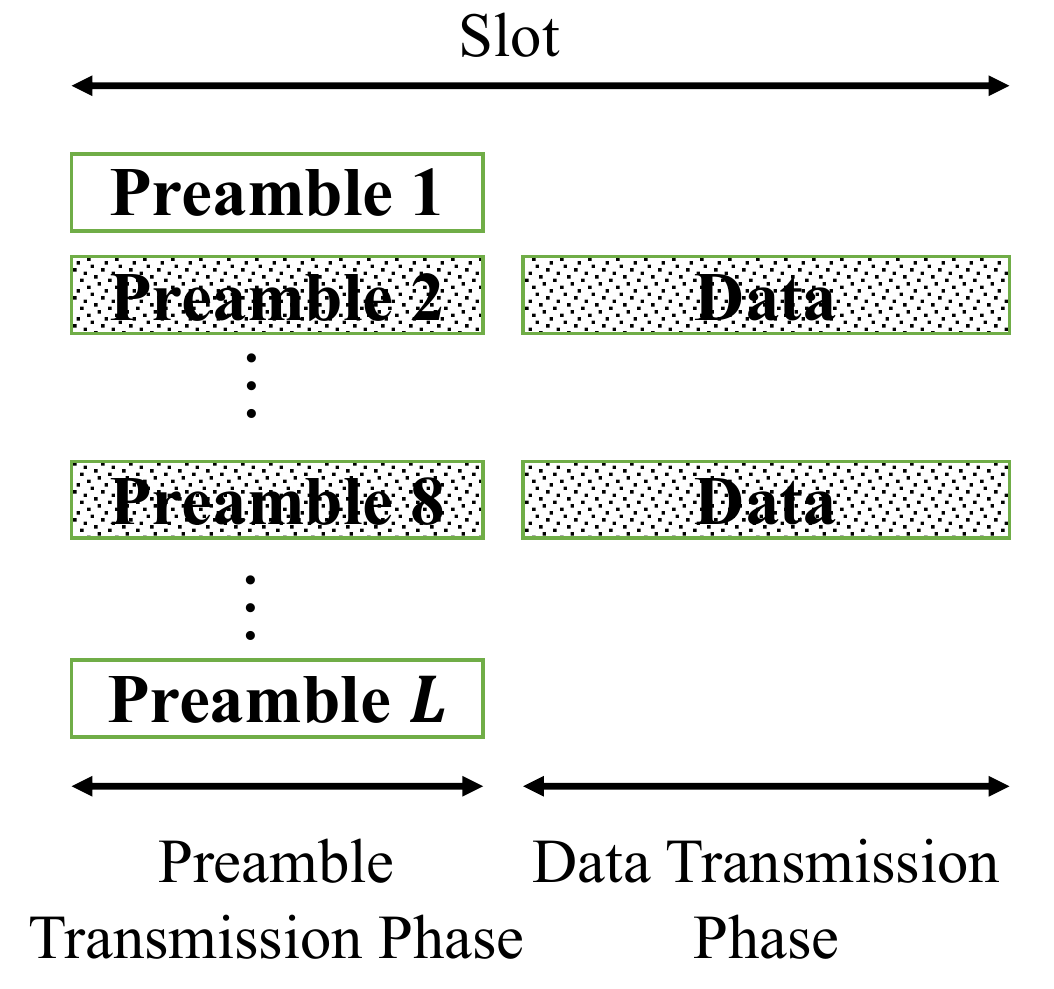}
\end{center}
\caption{A slot consisting of two sub-slots for
two different phases, namely
preamble and data transmission
phases, in a 2-step random access scheme (the shaded
blocks represent transmitted signals, i.e., there are
two active devices transmitting preambles 2 and 8).}
        \label{Fig:two_phase}
\end{figure}

Like the conventional 4-step random access, suppose that 
a pool of $L$ preambles is used 
in 2-step random access.
For data packet transmissions, 
a slot is divided into two sub-slots in a time division multiplexing (TDM)
manner. 
In the first sub-slot,  
each active device transmits a preamble that is chosen
from the preamble pool
uniformly at random and then transmits a data packet, which forms 
the first step, as illustrated in Fig~\ref{Fig:two_phase}.
In the second step, the receiver (which is 
a base station (BS)) sends the feedback signal to inform the decoding outcomes.

Suppose that there are $M$ active devices and consider an
active device
of interest that chooses a certain preamble. This device can 
successfully transmit its packet if the other devices
choose the other preambles,
and the corresponding probability is given by
\be 
p_s (M) =  \left(1 - \frac{1}{L} \right)^{M-1}.
\ee 
Thus, the probability of unsuccessful
packet transmission of the active device becomes
\begin{align}  
\epsilon & = \uE [1 - p_s (M) \,|\, M \ge 1]\cr
& = 1 - \uE \left[ \left(1 - \frac{1}{L} \right)^{M-1}
\,\bigl|\, M \ge 1 \right] \cr 
& = 1 - \sum_{m=1}^\infty \left(1 - \frac{1}{L} \right)^{m-1}
\Pr(M=m\,|\, M \ge 1),
    \label{EQ:ec2}
\end{align} 
where the expectation is carried out over $M$
and $\Pr(M=m\,|\, M \ge 1)$
is the conditional probability of $M =m$ provided that
$M \ge 1$.
That is, from an active device's point of view,
the channel can be seen as an erasure channel with
the erasure probability 
$\epsilon$ \cite{Choi21}. 
If $M$ follows a
Poisson distribution with mean $\lambda$,
the erasure probability becomes
\be 
\epsilon = 1 - \frac{e^{- \frac{\lambda}{L} }- e^{-\lambda}}{
(1 - e^{-\lambda}) \left(1 - \frac{1}{L}\right)}, \ L \ge 2,
\ee 
which can be approximated by $\frac{1}{L}$,
i.e., $\epsilon \approx \frac{1}{L}$, if $\lambda \ll L$.
In other words, in order to have a sufficiently low
$\epsilon$, say $10^{-5}$, $L$ has to be very large (i.e., $10^5$).

As shown in \eqref{EQ:ec1} and \eqref{EQ:ec2},
the channel erasure probability, $\epsilon$,
depends on a number of factors, and it would be difficult
to achieve a sufficiently low $\epsilon$ for ultra-reliable
communications. For example, as in \eqref{EQ:ec1},
the increase of the SNR, $\rho$, may not lead to the decrease
of $\epsilon$ if the difference between 
the channel capacity, $\log_2 (1+\rho)$ and the
code rate, $\frac{N_{\rm bit}}{n}$, is fixed
regardless of $\rho$.
To see this, let $\delta = \log_2 (1+\rho)- \frac{N_{\rm bit}}{n} > 0$,
which is assumed to be independent of $\rho$ and $n$.
As $\rho \to \infty$, $V(\rho) \to \bar V = (\log_2 e)^2$.
Thus, $\epsilon \to \cQ(\sqrt{\frac{n}{\bar V}} \delta)$ as $\rho \to
\infty$. In other words,
although the SNR, $\rho$, approaches infinity, $\epsilon$ cannot
approach 0, but a non-zero constant. 
In addition, from \eqref{EQ:ec2},
we can also see that $L$ should be sufficiently large
for a low erasure probability. However,
since the radio resource is limited, it is difficult to increase $L$.
As a result,
for ultra-reliable communications,
there should be diversity techniques, since the 
channel erasure probability, $\epsilon$ may not be sufficiently low.

\section{Reliable Transmissions}    \label{S:RT}

In this section, we discuss
reliable transmission of a message consisting of $M$ packets with a certain delay constraint for URLLC.

Throughout the paper, we have the following assumptions.
\begin{itemize}
    \item[{\bf A1}] As discussed in Section~\ref{S:SM},
each packet is independently transmitted
    over an erasure channel. 
    Since each packet is encoded with 
    parity bits,
    the receiver can decode it and declare its successful decoding 
or failure
    (then this packet is regarded as an erased one). As mentioned
    earlier, each packet is erased with
  a  probability of $\epsilon$.
\end{itemize}


\subsection{Repetition Diversity}

Suppose that each coded packet is transmitted $K$ times
according to $K$-repetition \cite{3GPP_HARQ1}.
This results in an improvement of
the reliability at the cost of the spectral efficiency
by a factor of $K$. For convenience, 
$\frac{1}{K}$ is referred to as the effective
spectral efficiency or 
code rate as $K$-repetition can be seen as a repetition code.

For convenience, assume that a block consists of $K$ slots.
Denote by $V_{k,m}$ the $k$th slot of block $m$.
Then,  with slight abuse of notation, let $V_{k,m} = X_m$
to represent that the packet transmitting in the $k$th slot of block $m$,
$k = 1, \ldots, K$, where $X_m$ represents original data packet $m$. 
That is, $K$ copies of $X_m$ is transmitted in a block
for $K$-repetition.

The receiver is able to decode the packet 
if at least one of $K$ copies 
can be correctly decoded.
Thus, according to 
Assumption of {\bf A1}, the decoding error probability becomes
\be
\epsilon_K = \binom{K}{0} \epsilon^K (1-\epsilon)^0 = \epsilon^K.
\ee
For example, if $\epsilon = 10^{-2}$,
in order to achieve a target error rate of $10^{-5}$,
$K$ should be greater than or equal to $3$.
The associated delay for each 
packet is the time duration of $K$ slots
or one block. 
In fact, this delay can be seen
as an upper-bound, because
one of the copies can be decoded before all $K$ copies are received.


In this paper, we consider the case that
a transmitter has a message of $M$ packets that 
are generated at a certain rate. When $K$-repetition
is employed, each packet can be successfully decoded
within a delay of $K$ slots with a probability 
of $\epsilon_K$. 
For a mission-critical application, the receiver may have to decode every packet within a certain delay, and decoding failures causing re-transmissions in HARQ 
could result in significant performance losses for the application.
Thus, when transmitting $M$ packets, it is desirable to have
a sufficiently low decoding error rate  to minimize
the number of re-transmissions.


\subsection{Random Linear Network Coding}   \label{SS:RLNC}

In this subsection, we consider an approach
that can effectively reduce the decoding error rate using NC
\cite{Choi_sub}.

As in \cite{Sundar08} \cite{Li11} \cite{Karzand17}, 
NC can be used for peer-to-peer communications.
In order to deliver $M$ packets, NC can be used.
Suppose that the transmitter uses 
random linear NC (RLNC) and the 
$n$th encoded packet is given by
\begin{align} 
Y_n & = f_n (X_1, \ldots, X_M) \cr
& = c_{n,1} X_1 \oplus \cdots \oplus c_{n,M} X_M, \ n = 1,\ldots, 
    \label{EQ:Yn}
\end{align} 
where $f_n(\cdot)$ is a random linear combination
of $M$ packets and 
the $c_{n,m}$'s are the encoding coefficients that are taken
from the Galois field, $GF(q)$. Here, $q$ represents
the size of the Galois field and
$\oplus$ represents the addition
in $GF(q)$, which is the XOR operation 
when $q = 2$. In \eqref{EQ:Yn},
the packet is also a vector over $GF(q)$.
Throughout the paper, 
a linear combination of data packets, i.e., $Y_n$ in \eqref{EQ:Yn}, 
is referred to as an NC packet, while $X_m$ is simply referred to as
a (data) packet.

Note that each NC packet is to be encoded
as an original data packet in $K$-repetition
so that each encoded packet
(whether it is an original packet, i.e., $X_m$, or
an NC packet, i.e., $Y_n$) is successfully received with
a probability of $1-\epsilon$ (or erased with a probability
of $\epsilon$)
according to Assumption of {\bf A1}.

Let $P (S,M)$ be the decoding probability
with $S\ (\ge M)$ successfully decoded NC packets. 
In \cite{Trull11}\cite{Zhao11}, with random encoding coefficients,
it is shown that
\be
P (S,M)
= \prod_{n=0}^{M-1} \left(1 - \frac{1}{q^{S-n}} \right).
    \label{EQ:Psm1}
\ee 
If all zero encoding coefficients are removed, 
$P_{\rm nc}$ is given by
\be 
P (S,M)
= 
\frac{\prod_{n=0}^{S-M} (-1)^n 
\binom{S}{n} U_q(M, S-n)  }{(q^M - 1)^S},
    \label{EQ:Psm2}
\ee 
where $U_q(m,n) = \prod_{j=0}^{m-1} (q^n - q^j)$. 
Note that since \eqref{EQ:Psm1} is a lower-bound on
\eqref{EQ:Psm2}, we will use \eqref{EQ:Psm1} to 
see the performance of NC in this paper.

\begin{figure}[thb]
\begin{center}
\includegraphics[width=\figwidth]{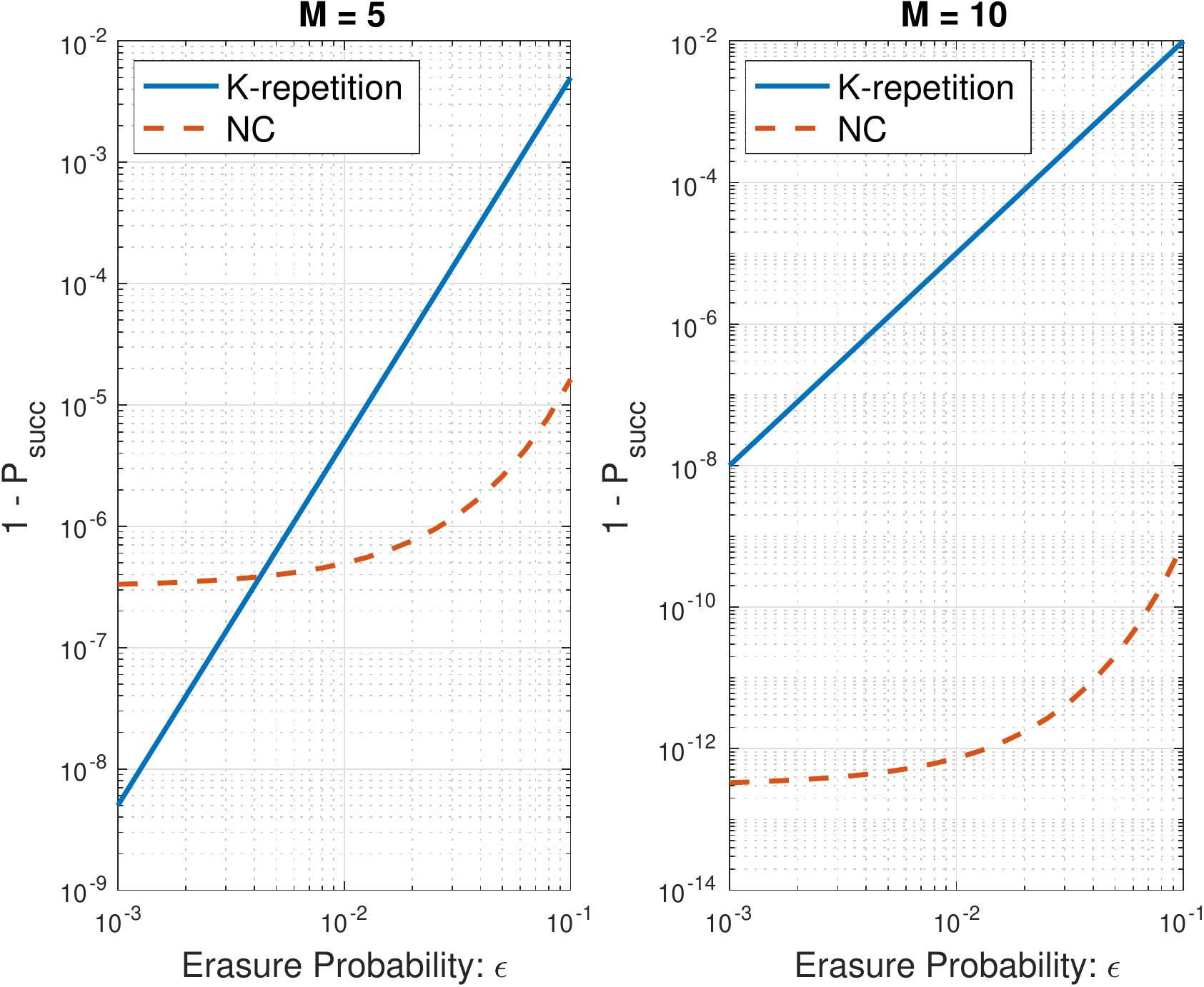} \\
\hskip 0.5cm (a) \hskip 3.5cm (b)
\end{center}
\caption{The error probability of $M$ packets,
i.e., $1 - P_{\rm nc} (N,M;\epsilon)$
and $1 - P_{\rm K} (N,M;\epsilon)$ for NC and $K$-repetition, respectively,
as functions of
channel erasure probability, $\epsilon$,
when $K = 3$ and $q = 4$ for NC: 
(a) $M = 5$;
(b) $M = 10$.}
        \label{Fig:plt_NCK}
\end{figure}

For a fair comparison with $K$-repetition
in terms of the spectral efficiency, suppose
that $KM$ NC packets are be transmitted.
With erasure probability $\epsilon$, the probability of successful
decoding of all $M$ packets is given by
\be
P_{\rm nc} (N,M; \epsilon) = \sum_{n = M}^{N} P (s, M) 
\binom{N}{s} (1-\epsilon)^s \epsilon^{N-s},
\ee 
where $N = MK$. For $K$-repetition,
the probability of successful
decoding of all $M$ packets becomes
\be
P_{\rm K} (N, M; \epsilon) = \left(1 - \epsilon_K \right)^M
= (1 - \epsilon^K)^M.
\ee 
In Fig.~\ref{Fig:plt_NCK}, 
the performance of NC and $K$-repetition 
is shown when $q = 4$, $K = 3$, and $M \in \{5, 10\}$.
Note that both NC and $K$-repetition transmit a total of $N = MK$ (coded) packets to transmit $M$ original data packets.
As $M$ increases, NC performs better than $K$-repetition. 
Note that the performance of $K$-repetition
is insensitive with respect to a finite $M$ if $\epsilon$ is
sufficiently low.

We have a few important remarks as follows.
\begin{itemize}
\item The use of NC can provide a significant improvement
in reliability compared to $K$-repetition
with the same spectral efficiency \cite{Choi_sub}.
However, since all $M$ packets are to be delivered, no specific
priority for the early packets is given. For example, in order
to decode the first packet, the receiver needs to wait to receive
at least $M$ NC packets.
This means that the lower bound on the delay of the first
packet is $M$. 
Thus, if $M > K$, the decoding delay of packet
of NC is longer than that of $K$-repetition. Note
that since $M$ is the lower bound on the delay in NC
and $K$ is the upper bound in $K$-repetition,
even if $M = K$, the packet transmission delay in NC
is expected to be longer than that in $K$-repetition.
Consequently, 
we can see that NC has a better reliability than $K$-repetition at
the cost of packet transmission delay, which means
that NC may not be suitable for URLLC.

\item As mentioned earlier, when the packets are generated
at a constant rate, the transmitter needs to wait till it
has $M$ original packets for NC. As a result, there would be an
additional delay at the transmitter side.

\item To reduce the decoding delay
in NC, a small number of packets can be considered.
For example, if there are 20 packets
to be delivered, we can divide them into 4 groups so that
each group has 5 packets. 
In this case, $M$ becomes 5 (not 20). However, as shown in
Fig.~\ref{Fig:plt_NCK} (a), if $\epsilon$ is sufficiently low
(say $\epsilon = 10^{-3}$), the 
decoding error probability of NC can be higher than
that of $K$-repetition (in addition to this,
the decoding
delay, which is $M = 5$, is longer than that of $K$-repetition, which is $K = 3$). This (i.e., the use of small
number of packets for NC) offsets the performance gain of NC.
\end{itemize}

\section{Sliding Network Coding}    \label{S:SNC}

In this section, we introduce
SNC that can provide a relatively low transmission
delay 
for each packet with improved reliability
compared to $K$-repetition.
In particular, SNC can take advantage
of both $K$-repetition and NC using on-the-fly mode.


\subsection{Examples}

In order to illustrate the idea of SNC, consider an example with $K = 2$.
Suppose that a block consists of two consecutive slots,
which is denoted by $(V_{1,m}, V_{2,m})$. Here,
$V_{k,m}$ is the NC packet transmitted in the $k$th slot of block $m$.
As shown in Table~\ref{TBL:SNC1},
for example, the NC packets are given.
For $m \ge 2$, we have
\be
(V_{1,m}, V_{2,m}) = (X_m, X_{m-1} \oplus X_m).
\ee 
\begin{table}[h]
     \caption{An example of SNC with $K = 2$ and a delay
     of 1 block.}
     \centering
    \begin{tabular}{c||c|c|c|c|c}
     $m$ & 1 & 2 & 3 & 4& $\cdots$\\ \hline
      $V_{1,m}$ & $X_1$  & $X_2$ & $X_3$ &  $X_4$ & $\cdots$\\
    $V_{2,m}$   &  $X_1$ & $X_1 \oplus X_2$ & $X_2 \oplus X_3$ & $X_3 \oplus X_4$ & $\cdots$ \\
    \end{tabular}
    \label{TBL:SNC1}
\end{table}

At the end of block $m$, suppose that
the receiver is to decode
$X_{m-1}$. For example, consider $m = 4$ and  the
receiver is to decode $X_3$.
Then, we have $V_{1,4} = X_4$ and $V_{2,4} = X_3 \oplus X_4$.
In addition, the receiver also has $V_{1,3} = X_3$ and $V_{2,3} = X_2
\oplus X_3$. 
Suppose that all the previous packets are successfully decoded.
This means that $X_1$ and $X_2$ are decoded.
Then, according to Assumption of {\bf A1}, the decoding error 
probability of $X_3$
is as follows:
\begin{align} 
P_3 & = \underbrace{\epsilon^2}_{(a)}
\underbrace{(1 - (1-\epsilon)^2)}_{(b)} \cr
& = 2\epsilon^3 - \epsilon^4. 
    \label{EQ:P3}
\end{align}  
The part (a) is due to the
fact that the decoding error probability when the receiver
receives $V_{1,3} = X_3$ and that when $V_{2,3} = X_2 \oplus X_3$
(since $X_2$ is assumed to be correctly decoded, 
$X_3$ can be decoded if $X_2 \oplus X_3$ is correctly decoded,
which means that the decoding error probability
with $X_2 \oplus X_3$ is $\epsilon$).
The part (b) is due to the decoding error when $X_3$ is to 
decoded with
$V_{2,4} = X_3 \oplus X_4$ and $V_{1,4} = X_4$.
For successful decoding of $X_3$,  both $V_{2,4}$ and $V_{1,4}$ 
should be correctly decoded.
Thus, the associated error probability of decoding is $1 - (1-\epsilon)^2$.
It can also be shown that $X_1$ is decoded at $m = 2$
with the following decoding error probability: 
\be 
P_1 = \epsilon^2 ( 1 - (1-\epsilon)^2) = P_m = 2 \epsilon^3+O(\epsilon^4),
\ee 
for $m \ge 2$.

In order to see the advantage of SNC over $K$-repetition,
let consider an example with $\epsilon = 0.01$ and 
a target decoding
error probability of $p_{\rm err} = 10^{-5}$.
When
$K$-repetition is used, we need to have  $K \ge 3$.
On the other hand, with SNC, as shown above, $K = 2$ is sufficient
as the decoding error probability becomes
$2 \epsilon^3 = 2 \times 10^{-6}$.
That is, with a small number of slots per block (or repetitions),
a lower decoding error probability
can be achieved using SNC.

For decoding delay, compared to $K$-repetition, 
SNC has an additional delay of one  block
as $X_{m-1}$ is to be decoded at the end of block $m$.
Note that 
the resulting approach is referred to as SNC,
because
$V_{2,m}$, which is an XOR of two packets, $X_{m-1}$
and $X_m$, is a linear combination of the packets
within a sliding window of two consecutive packets
(this becomes clear with $K \ge 2$, which will be discussed
later).

Note that any incorrect decoding of the data packets 
will result in subsequent decoding errors, i.e., 
there is error propagation in SNC.
Thus, it may be necessary to lower the decoding error
probability of the first data packet, $X_1$.
To this end, consider the example in Table~\ref{TBL:SNC2}.
It can be readily shown that $X_1$ can be decoded at $m = 3$
with the following decoding error probability: 
\be 
P_1 = \epsilon^2 ( 1 - (1-\epsilon)^2)^2 = 4 \epsilon^4+O(\epsilon^5).
\ee 
Clearly, we have this decrease of the decoding error probability of $X_1$
at the cost of delay. That is, $X_{m-2}$ can be decoded at block $m$.

\begin{table}[h]
     \caption{An example of SNC with $K = 2$ and a delay of 2 blocks.}
     \centering
    \begin{tabular}{c||c|c|c|c|c}
     $m$ & 1 & 2 & 3 & 4& $\cdots$\\ \hline
     $V_{1,m}$ & $X_1$  & $X_2$ & $X_3$ &  $X_4$ & $\cdots$\\
     $V_{2,m}$ & $X_1$ & $X_1 \oplus X_2$ & $X_1 \oplus X_3$ & $X_2 \oplus X_4$ & $\cdots$ \\
    \end{tabular}
    \label{TBL:SNC2}
\end{table}

We can see that the two designs 
in Tables~\ref{TBL:SNC1}
and~\ref{TBL:SNC2} have
a spectral efficiency of $\frac{1}{K} = \frac{1}{2}$ and 
a decoding error rate of $O(\epsilon^3)$.
While the resulting decoding error can be sufficiently
low for some applications,
it is also possible to have further lower error rates
with $K \ge 2$ via a generalization of SNC,
which will be discussed in the next subsection.

\subsection{A Design with More than Two Slots per Block}
\label{SS:K}

In this subsection, we consider SNC with $K \ge 2$
through a generalization.

As mentioned earlier, 
we assume that a block consists of $K$ slots. 
Denote by $D$ the delay parameter such that
$X_{m-D}$ is to be decoded at the end of block $m$.
The encoded packets of block $m$ are now given by
\begin{align}
V_{1,m} &= X_m \cr
V_{k,m} &= X_{m-D} \oplus f_k (X_m, \ldots, X_{m-D+1}),
k = 2,\ldots, K, \quad
    \label{EQ:gen}
\end{align}
where the $f_k (\cdot)$ are different linear combinations
of the data packets, $X_m, \ldots, X_{m-D+1}$, at block $m$.
That is, the $(K-1)$ NC packets are 
\begin{align}
& f_k(X_m, \ldots, X_{m-D+1}) = \cr
& \quad c_{k,1} X_{m} \oplus \cdots \oplus c_{k,D} X_{m-D+1},
	\label{EQ:fk}
\end{align}
where $c_{k,d} \in GF (q)$ is the encoding coefficient for 
the $m$th slot, $V_{k,m}$, or 
\be
\left[ \begin{array}{c} 
f_1 \cr
\vdots  \cr
f_{K-1} \cr
\end{array}
\right] = \bC 
\left[ \begin{array}{c} 
X_m \cr
\vdots  \cr
X_{m-D+1} \cr
\end{array}
\right],
    \label{EQ:fCX}
\ee 
where $[\bC]_{k,d} = c_{k,d}$ and the size of $\bC$ is $(K-1) \times D$.
For convenience, the SNC in \eqref{EQ:gen}
is referred to the $(K,D,q)$-SNC design,
where $K$ represents the inverse of the 
effective spectral efficiency, $D$ represents the delay in block,
and $q$ is the size of Galois field.
Consequently, we can see that each block $m$ consists
of the current original data packet, $X_m$, and $K-1$ NC packets
that are linear combinations of the current and past packets
in SNC.

Note that 
unlike the approach of NC in Subsection~\ref{SS:RLNC},
since the packets in block $m$ are linear combinations
of current and past packets, 
the transmitter has no encoding delay to form NC packets, 
and is 
able to send a new original packet per block in on-the-fly mode.

\begin{figure}[thb]
\begin{center}
\includegraphics[width=\figwidth]{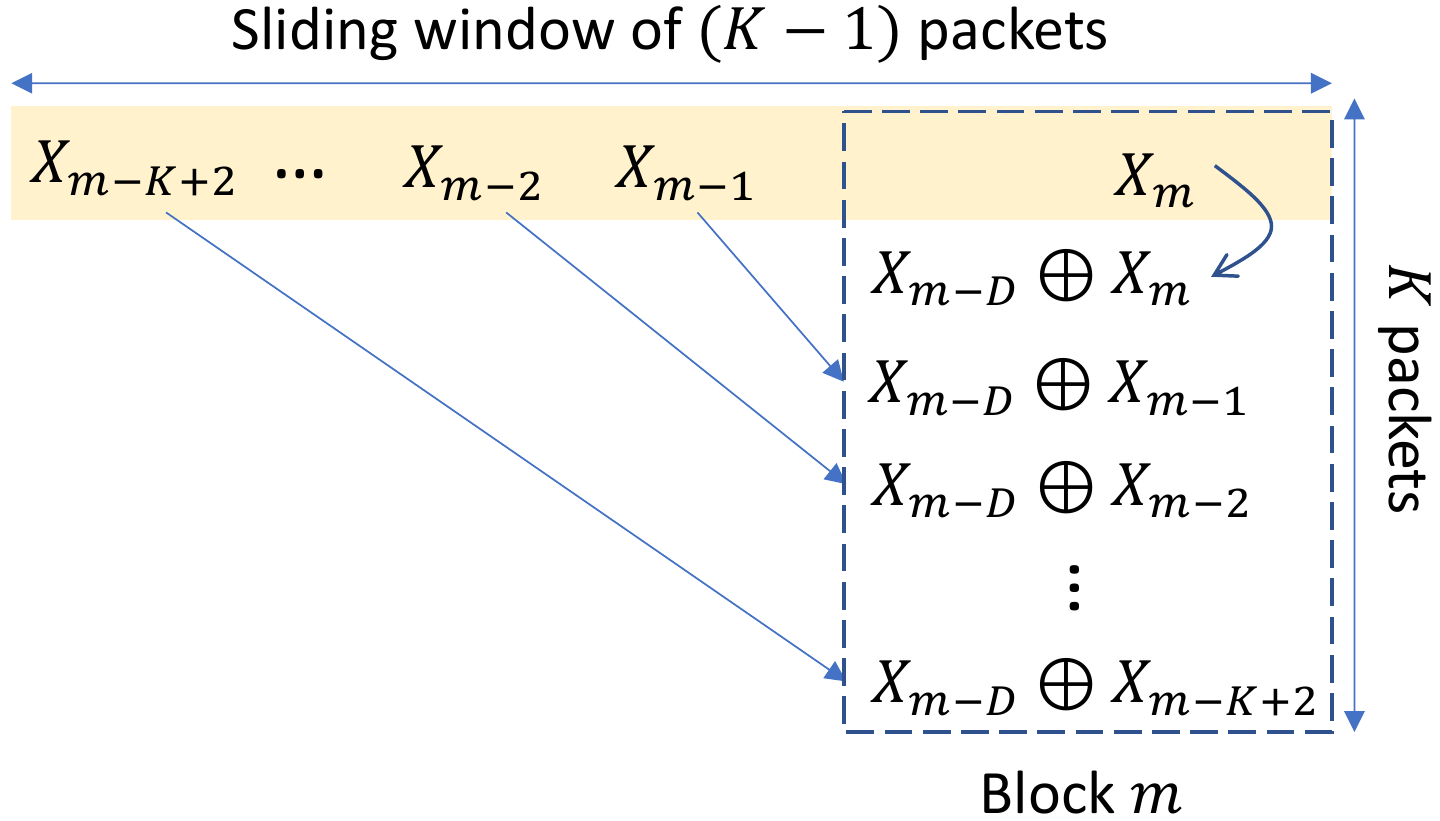}
\end{center}
\caption{An illustration to show the 
sliding window to generate NC packets in block $m$,
where $D = K-1$.}
        \label{Fig:snc}
\end{figure}

For example, 
consider the following simple design:
\begin{align}
    V_{1,m} & = X_m \cr
    V_{k,m} & = X_{m-D} \oplus X_{m-k+2}, \ k = 2,\ldots,K,
    \label{EQ:VKs}
\end{align}
where linear combinations of two packets are considered for
the NC packets, $V_{k,m}$, $k = 2, \ldots, K$.
In this design, the delay parameter, $D$, becomes $K-1$
and $\bC = \bI$.
In Fig.~\ref{Fig:snc}, a sliding window 
of $K-1$ packets is shown to form 
$K-1$ NC packets in block $m$.
From this, it is clear
that $D$ becomes $K-1$. 
That is, \eqref{EQ:VKs} is a $(K,K-1,2)$-SNC.

While the $(K,K-1,2)$-SNC design in \eqref{EQ:VKs}
is simple (as only two packets are combined for NC packets),
the decoding delay can be long for a large $K$,
In particular, the decoding delay becomes
$K^2$ (in slots), which is $K$-time
longer than that of $K$-repetition.
This shows that SNC can provide a higher reliability at the cost of delay.
Since it is desirable to have a short decoding delay,
we now find the minimum decoding
delay, $D$, for $(K,D,q)$-SNC.

\begin{mylemma} \label{L:Delay}
For $(K,D,q)$-SNC,
the delay parameter, $D$, has to satisfy the following inequality:
\be 
D \ge \log_q K.
    \label{EQ:DK}
\ee 
\end{mylemma}
\begin{IEEEproof} 
For a given $q$, from \eqref{EQ:fk},
there can be $q^D - 1$ non-zero different NC packets
that are linear combinations of $X_{m}, \ldots, X_{m-D+1}$.
Thus, we have $K -1 \le q^D - 1$ or $K \le q^D$,
which leads to \eqref{EQ:DK}.
\end{IEEEproof}

According to \eqref{EQ:DK},
we can see that the minimum decoding delay
increases logarithmically with $K$.
Any $(K,D,q)$-SNC design with $D = \lceil \log_q K \rceil$ is referred to 
as a minimum delay $(K,D,q)$-SNC.
Clearly,
the SNC in Table~\ref{TBL:SNC1} is
an example of a minimum delay $(K,D,q)$-SNC design,
where $(K,D,q) = (2,1,2)$, while
the SNC in Table~\ref{TBL:SNC2} with $(K,D,q) = (2,2,2)$
does not have the minimum delay.
Another example of a minimum delay $(K,D,q)$-SNC design
with $(K,D,q) = (4,2,2)$ 
can also be found in Table~\ref{TBL:SNC_g}.

\begin{table*}[ht]
     \caption{NC packets of SNC with $q=2$, $K = 4$, and $D = 2$.}
     \centering
    \begin{tabular}{c||l|l|l} \hline
block & $m-2$ & $m-1$ & $m$ \\ \hline
     $V_{1,.}$ &  $X_{m-2}$ & $X_{m-1}$ &  $X_m$ \\
     $V_{2,.}$  & $X_{m-4} \oplus X_{m-2}$ & $X_{m-3} \oplus X_{m-1}$ & $X_{m-2} \oplus X_m$  \\
     $V_{3,.}$   & $X_{m-4} \oplus X_{m-3}$ & $X_{m-3} \oplus X_{m-2}$ & $X_{m-2} \oplus X_{m-1}$  \\
     $V_{4,.}$   & $X_{m-4} \oplus X_{m-2} \oplus X_{m-3}$ & $X_{m-3} \oplus X_{m-1}
     \oplus X_{m-2}$ & $X_{m-2} \oplus X_m \oplus X_{m-1}$  \\ \hline
    \end{tabular}
    \label{TBL:SNC_g}
\end{table*}

\subsection{A Decoding Rule}


At the receiver, noisy versions of $V_{k,m}$, which
are denoted by $\hat V_{k,m}$, are received
through an erasure channel. Under the assumption of {\bf A1},
we have
\be
\hat V_{k,m} = \left\{
\begin{array}{ll} 
V_{k,m} & \mbox{w.p. $1 - \epsilon$} \cr 
? & \mbox{w.p. $\epsilon$,} \cr 
\end{array}
\right.
\ee 
where $?$ represents the erasure.
For decoding, we consider the following two stages:
\begin{enumerate}
    \item[S1] At the end of block $m$,
    the received NC packets, $\hat V_{k,m}$, $k = 1,\ldots, K$,
    are individually decoded. They are referred to as
    not-fully-decoded (NFD) packets.
Since the channel is an erasure channel, each NFD packet is
 either successfully decoded (with probability
 $1 - \epsilon$) or unknown (with probability of $\epsilon$).
    \item[S2] Then, $X_{m-D}$ is to be decoded
    using the NFD packets and the fully-decoded (FD) packets of 
    $X_{m-D-k}$, $k \ge 1$.
\end{enumerate}
In the second step, since $X_{m-D}$ is to be decoded at the end of
slot $m$,  $X_{m-D-k}$, $k \ge 1$, 
should be decoded in the previous blocks.
Thus, they are available as FD packets. 
That is, at the end of block $m$, we have
\be
\underbrace{\ldots, X_{m-D-1}}_{\rm FD\ packets}, X_{m-D},
\underbrace{X_{m-D+1},\ldots, X_{m}}_{\rm NFD\ packets} .
\ee

To illustrate the decoding rule, consider an example with
the SNC design in Table~\ref{TBL:SNC_g}.
At the end of block $m$, suppose that the 
receiver finds that the following NFC packets are erased after
the first step:
\begin{align}
    \hat V_{1,m} = \hat V_{3,m} = \hat V_{4,m-1} = \hat V_{1,m-2} = 
    \hat V_{2,m-2} = ?,
\end{align}
while the other NFD packets and all the FD packets are correctly decoded.
Then, in the second step, the receiver has the following NC packets directly
related to $X_{m-2}$:
\begin{align}
    V_{2,m} & = X_{m-2} \oplus X_m \cr
    V_{4,m} & = X_{m-2} \oplus X_m \oplus X_{m-1} \cr 
    V_{3,m-1} & = \fbox{$X_{m-3}$} \oplus X_{m-2} \cr 
    V_{4,m-2} & = \fbox{$X_{m-4}$} \oplus \fbox{$X_{m-3}$} \oplus X_{m-2},
\end{align}
where the boxed variables are FD packets, which are assumed to be
correct. Then, we can see that $X_{m-2}$
can be decoded from $X_{3,m-1}$ or $X_{4,m-2}$.
Note that once $X_{m-2}$ is successfully decoded,
it becomes an FD packet, which can help decode
$X_m$ from $V_{2,m}$.

\section{Performance Analysis}  \label{S:PA}

In this section, 
we present the performance analysis of SNC
in terms of decoding error rate.

First, we consider the
simple design in \eqref{EQ:VKs}, i.e., $(K,K-1,q)$-SNC,
which allows a tractable analysis
to find the decoding error rate. As shown below,
it seems that the error exponent of SNC
can be about two times higher than that of $K$-repetition.

\begin{mylemma} \label{L:1}
Suppose that the receiver is to decode
$X_{m-D}$ at the end of block $m$
when the SNC in \eqref{EQ:VKs} is used with $D = K-1$.
Provided that all the FD packets
are correctly decoded, 
the decoding error probability
of $X_{m-D}$ is given by
\begin{align}  
p_{\rm snc} & = \epsilon^K (1 - (1-\epsilon)^2)^{K-1} \cr
& = 2^{K-1} \epsilon^{2K-1}
+O(\epsilon^{2K}).
    \label{EQ:L1}
\end{align} 
\end{mylemma}
\begin{IEEEproof}
To find the decoding error probability of $X_{m-D}$,
two different sets of received signals are considered.

\begin{enumerate}
    \item With the current block $m$, 
there are $(K-1)$ (XORed) copies of $X_{m-D}$ with the NFD packets,
i.e., $V_{k,m}$, $k = 2,\ldots, D$. 
The decoding error becomes $(1-(1-\epsilon)^2)^{K-1}$.

\item  There are $K$ past blocks
that contain copies of $X_{m-D} = X_{m-K+1}$. For example,
at block $m-K+1$,
$V_{1,m-K+1} = X_{m-K+1}$ was transmitted according to \eqref{EQ:VKs}.
In addition, 
at block $m-d$, $V_{K-d+1,m-d} = X_{m-d-K+1} \oplus X_{m-K+1}$,
$d =1, \ldots, K-1$, is transmitted.
Since we assumed that all the FD packets are correctly decoded,
in $V_{K-d+1,m-d} = X_{m-d-K+1} \oplus X_{m-K+1}$,
$X_{m-d-K+1}$ is known if $V_{k-d+1,m-d}$ is correctly
decoded. Thus, 
the associated error probability to decode $X_{m-K+1}$
for given $V_{K-d+1,m-d}$ is $\epsilon$.
As a result, 
the error probability with the received signals in the past blocks,
i.e.,
$V_{1,m-K+1}$ and $V_{K-d+1,m-d}$,
$d \in \{1, \ldots, K-1\}$,
becomes $\epsilon^K$.
\end{enumerate}

Consequently, the decoding error probability
becomes the product of $\epsilon^K$
and $(1-(1-\epsilon)^2)^{K-1}$, which is given in \eqref{EQ:L1}.
\end{IEEEproof}

There are some remarks.
\begin{itemize}
    \item The SNC in Table~\ref{TBL:SNC1} is a $(K,K-1,2)$-SNC design
with $K = 2$. As shown in \eqref{EQ:L1},
the decoding error rate is 
$p_{\rm nsc} = 2 \epsilon^3 + O(\epsilon^4)$,
which agrees with that in \eqref{EQ:P3}.

\item The decoding error rate in \eqref{EQ:L1}
can be regarded as an upper-bound as it is assumed that
the erasure probability of
any NFD packet in block $m$,
$X_{m},\ldots, X_{m-D+1}$, is set to $\epsilon$.
In practice, some of the NFD packets can be decoded in the
previous decoding rounds. Thus, the effective erasure probability
can be lower than $\epsilon$, which results
in the actual decoding error rate
that is  lower than that in \eqref{EQ:L1}. 


\end{itemize}

As shown in \eqref{EQ:L1}, an approximate decoding error rate of 
the $(K,K-1,2)$-SNC design in \eqref{EQ:VKs}
is available. However, in a general $(K,D,q)$-SNC design,
it is not straightforward to find such an expression and
we need to define a few more parameters.

Define $\mu$ as the number of NCs in blocks $m-D, \ldots, m-1$ 
that are $X_{m-D}$ itself or linear combinations of $X_{m-D}$
and FD packets, $X_{m-D-k}$, $k \ge 1$.
As an example, consider the SNC design in Table~\ref{TBL:SNC_g}.
Since $X_{m-3}$ and $X_{m-4}$
are FC packets, it can be shown that
\begin{align*}
    V_{1,m-2} &= X_{m-2} \cr
V_{2,m-2} &= X_{m-4} \oplus X_{m-2} \cr 
V_{4,m-2} &= X_{m-4} \oplus X_{m-3} \oplus X_{m-2} \cr
V_{3,m-1} &= X_{m-3} \oplus X_{m-2}.
\end{align*}
Thus, we have $\mu = 4$ in this example. 

\begin{mylemma}     \label{L:3}
Suppose that $D \le K - 1$ and
the coefficient matrix $\bC$ in
\eqref{EQ:fCX} can be expressed (possibly after permutation of rows) as
\be
\bC = \left[ \begin{array}{c} 
{\rm diag}(c_{1,1}, \ldots, c_{D,D} ) \cr \bar \bC \cr
\end{array} \right],
    \label{EQ:Lcond}
\ee 
where $c_{d,d} \ne 0$, $d = 1,\ldots, D$.
The decoding error rate of a
$(K,D,q)$-SNC design is given by
\be
p_{\rm snc} = 2^D \epsilon^{h} + O(\epsilon^{h+1}), \ h \ge \mu + D,
    \label{EQ:L3}
\ee 
where $\mu \ge D$.
\end{mylemma} 
\begin{IEEEproof}
Eq. \eqref{EQ:Lcond} implies that
there exist $D$ $f_k$'s such that
\be 
f_k = c_{k,d} X_{m-d}, \ c_{k,d} \ne 0, \ k \in \{k_1, \ldots, k_D\}. 
\ee 
At the end of block $m$, the receiver
has $\hat V_{1,m-d}$, 
which are noisy versions of NFD packets, $X_{m-d}$, $d = 0,\ldots, D-1$,
as shown in \eqref{EQ:gen}. Thus, to decode $X_{m-D}$
from a pair of $(\hat V_{1,m-d}, \hat V_{k,m})$, $k \in \{k_1,
\ldots, K_D\}$, they should not be erased. The corresponding
probability is $(1-\epsilon)^2$. Since there are $D$ pairs,
the decoding error probability from NFD packets
is $(1 - (1-\epsilon)^2)^D = (2 \epsilon)^D + O(\epsilon^{D+1})$.
The receiver can also decode $X_{m-D}$ using FD packets.
Thus, by
the definition of $\mu$, we
can see that the exponent
of the decoding error rate with FD packets can be
greater than or equal to $\mu$.
As a result, the decoding error rate becomes $\epsilon^n
\left((2 \epsilon)^D
+ O(\epsilon^{D+1}) \right) = 2^D \epsilon^{n+D} +
O(\epsilon^{n+ D+1})$, where 
$n \ge \mu$. 

Thanks to the structure of $\bC$ in \eqref{EQ:Lcond},
there are at least $D$ NCs that include $X_{m-D}$
(as $X_{m-D-d} \oplus X_{m-D}$
can be found in block $m-d$, $d = 1,\ldots,D$). Thus, $\mu \ge D$.
This completes the proof.
\end{IEEEproof}

Using Lemma~\ref{L:3}, we can find an approximate
decoding error 
rate of any $(K,D,q)$-SNC design.
For example, the SNC design
in Table~\ref{TBL:SNC_g} has a decoding error rate of
 $(2 \epsilon)^D \epsilon^\mu = 2^D \epsilon^{D+\mu} =
4 \epsilon^6$, which can be seen as an upper-bound.

We have a few remarks.
\begin{itemize}
    \item As shown in \eqref{EQ:L3}, the 
    decoding error rate of SNC can decrease with the delay parameter $D$.
    That is, there is a trade-off between the delay and reliability.
    \item  As shown 
   in Lemma~\ref{L:3}, if $\bC$ is designed as in \eqref{EQ:Lcond},
   we expect that $\mu$ increases with $D$
   (since $\mu$ is upper-bounded by $D$). However,
there might be a better design to maximize $\mu$ 
or minimize the decoding error rate for given $K$ and $D$. 
Finding an optimal SNC design is a further research topic
to be studied in the future.
\end{itemize}

\section{Simulation Results}    \label{S:Sim}

In this section, we present
simulation results with two 
different SNC designs,
$(3,2,2)$-SNC and $(4,2,2)$-SNC, unless stated otherwise,
and compare them with those of $K$-repetition.
For simplicity, we only consider the case of $q = 2$
(i.e., binary NC). 

\begin{figure}[thb]
\begin{center}
\includegraphics[width=\figwidth]{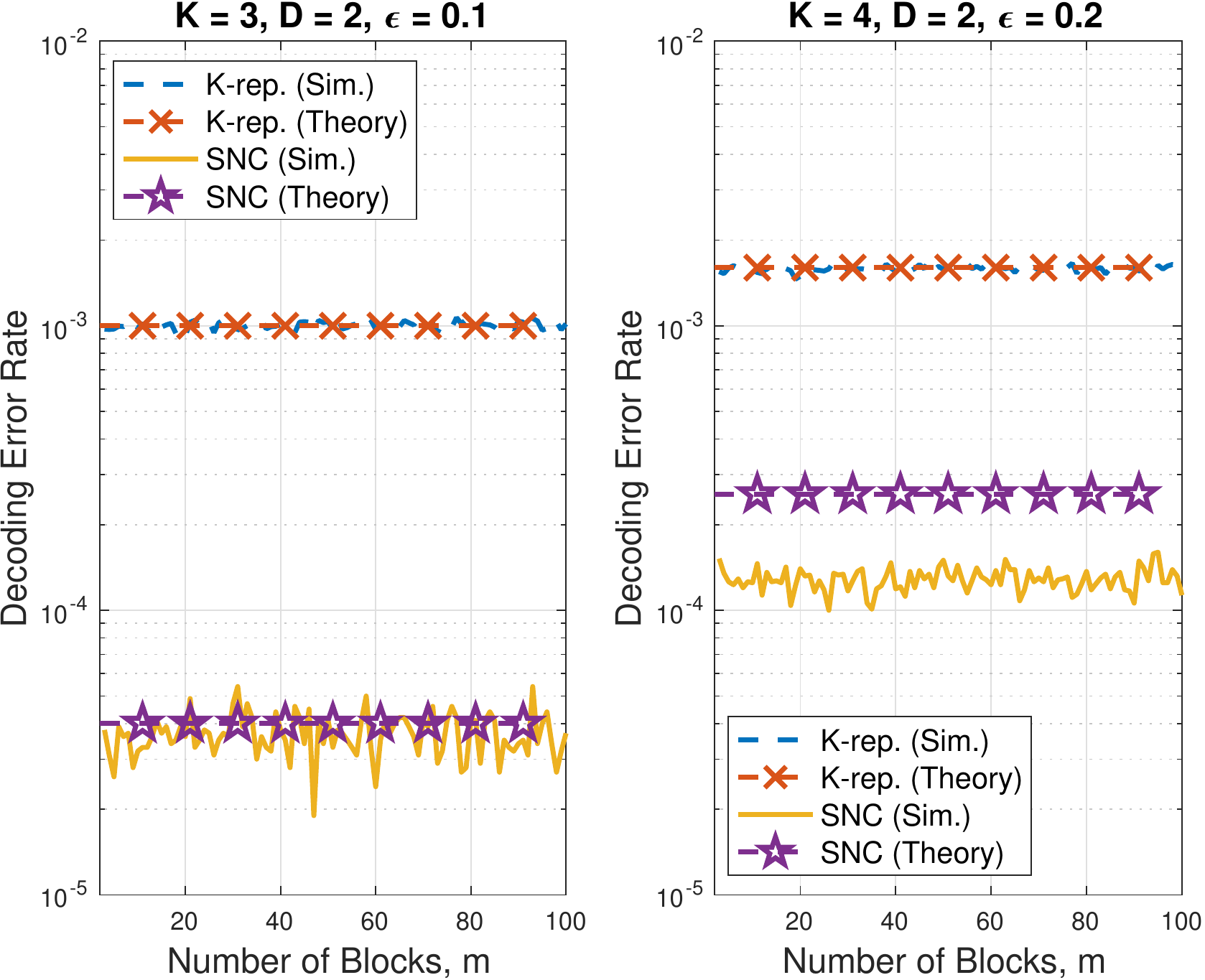} \\
\hskip 0.5cm (a) \hskip 3.5cm (b)
\end{center}
\caption{Decoding error rates of 
SNC and $K$-repetition (bounds for SNC):
(a) $K = 3$, $D = 2$, and $\epsilon = 0.1$;
(b) $K = 4$, $D = 2$, and $\epsilon = 0.2$.}
        \label{Fig:plt_KDs}
\end{figure}

In Fig.~\ref{Fig:plt_KDs}, 
the decoding error rate is shown over time
(in blocks). 
As shown in Fig.~\ref{Fig:plt_KDs} (a)
for the performance when $K = 3$ and $\epsilon = 0.1$,
$K$-repetition has a decoding
error rate of $\epsilon^3 = 10^{-3}$,
while SNC provides a much lower error rate,
which is about $4 \epsilon^5 = 4 \times 10^{-5}$
(according to \eqref{EQ:L1}).
That is, at the same spectral efficiency,
SNC can provide a much lower decoding error
rate than $K$-repetition at the cost of additional
decoding delay.
Fig.~\ref{Fig:plt_KDs} (b) shows
the performance when $K = 4$ and $\epsilon = 0.2$.
We can see that the $(4,2,2)$-SNC
design can 
perform better than $K$-repetition as expected.
From Lemma~\ref{L:3}, we can find an upper-bound
on the decoding error probability of 
$(4,2,2)$-SNC, which is $4 \epsilon^6 \approx 2.56\times 10^{-4}$.
As shown in
Fig.~\ref{Fig:plt_KDs} (b), indeed, it is an upper-bound. 
Note that the decoding error rate 
of $(4,3,2)$-SNC can be
$2^3 \epsilon^{7} 
\approx 1.02 \times 10^{-4}$ according to \eqref{EQ:L1},
which is close to the actual decoding error rate of 
$(4,2,2)$-SNC. Thus,
the decoding delay can be less than 
$D = K-1$ in blocks 
without significant performance
degradation in terms of decoding error rate,
if SNC can be carefully designed.

\begin{figure}[thb]
\begin{center}
\includegraphics[width=\figwidth]{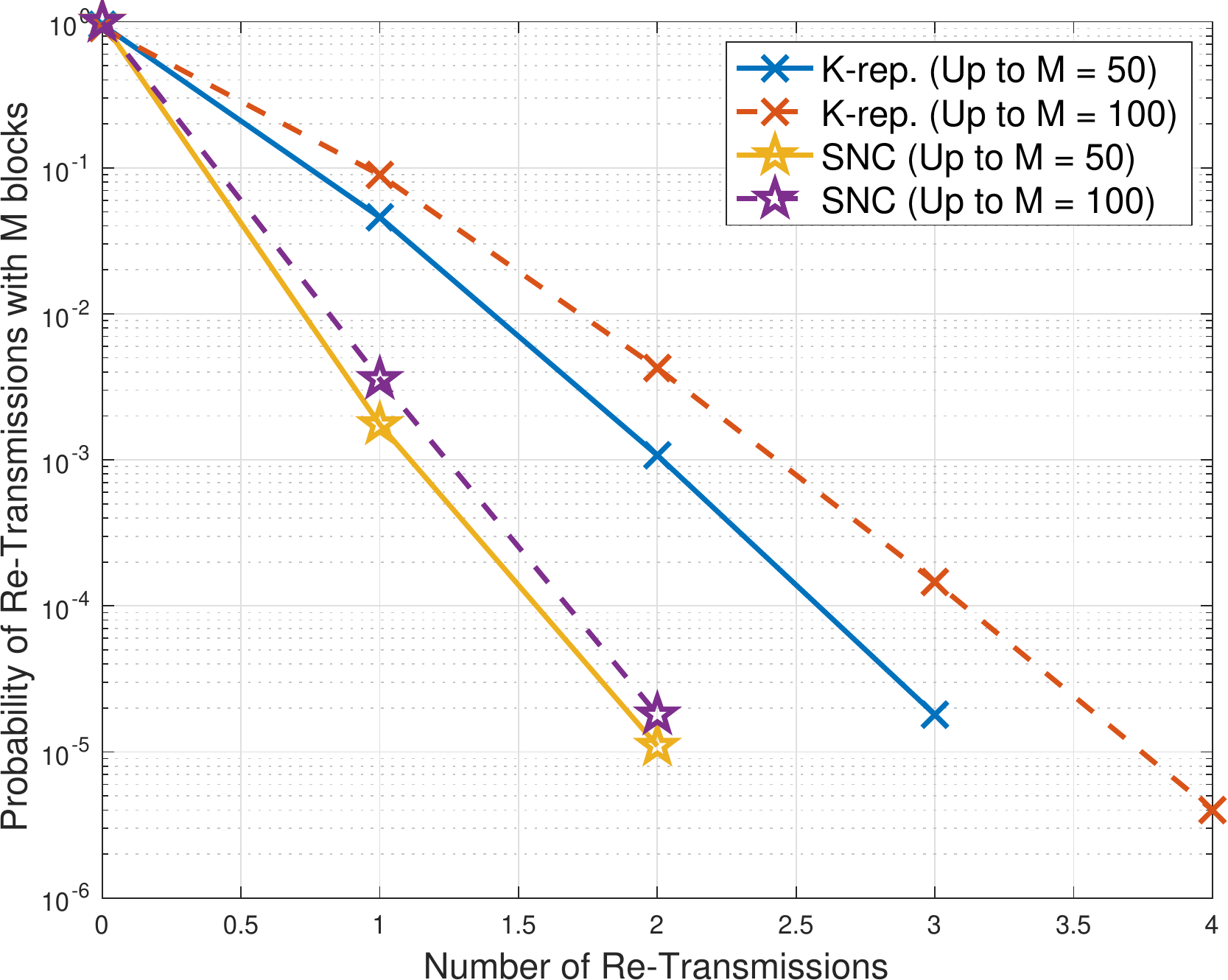} \\
(a) \\
\includegraphics[width=\figwidth]{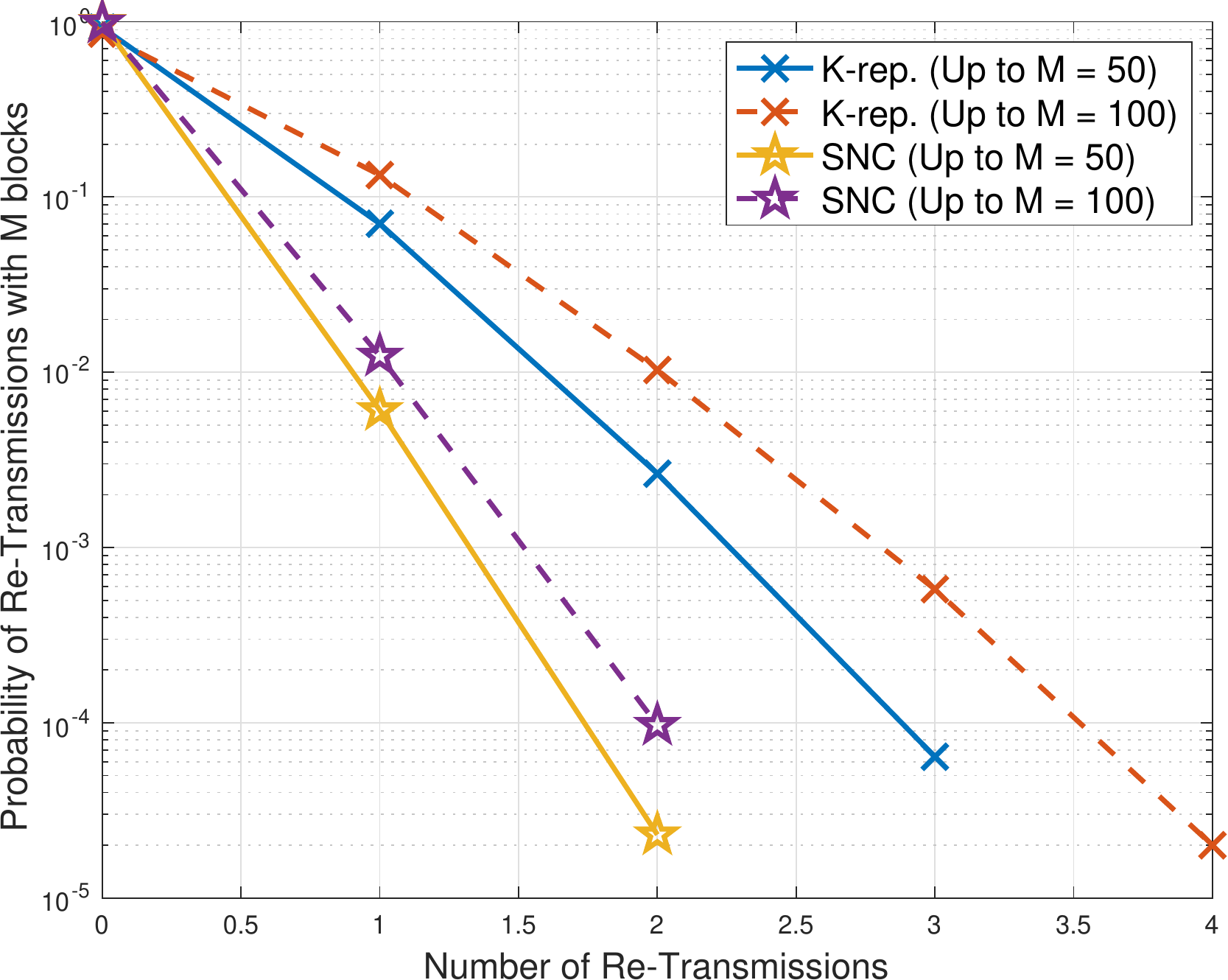} \\
(b) 
\end{center}
\caption{Probability of re-transmissions 
versus number of re-transmissions within a session of
$M \in \{50, 100\}$ blocks:
(a) $K = 3$, $D = 2$, and $\epsilon = 0.1$;
(b) $K = 4$, $D = 2$, and $\epsilon = 0.2$.}
        \label{Fig:plt_rtx}
\end{figure}

Although $K$-repetition
or SNC can lower the decoding error rate, it is impossible to completely avoid decoding failures. 
Thus, we may need to re-transmit packets if necessary in some applications.
In Fig.~\ref{Fig:plt_rtx},
the probability of a certain number of re-transmissions 
is shown 
when a user transmits a message consisting
of $M \in \{50,100\}$ packets.
That is, assuming that a user is to transmit
a message of $M$ packets over a session,
we obtain
the probability that
the total number of re-transmissions of packets 
is $i \in \{0,\ldots\}$ within a session.
As shown in Fig.~\ref{Fig:plt_rtx} (a),
the probability that one of $M= 100$ packets is to be re-transmitted
is about $10^{-1}$ with $K$-repetition, while
this probability becomes less than $4 \times 10^{-3}$
with SNC.
Clearly, SNC can significantly reduce the number of re-transmissions,
which is important for URLLC design as
each re-transmission results in additional packet 
transmission delay. In some mission-critical applications, additional unexpected delays can result in significant performance losses.
Thus, for such mission-critical applications, SNC
can be a good candidate as it can provide a high reliability
with a guaranteed delay, i.e., a very low decoding error rate,
say $O(\epsilon^{\mu+D})$,
with a delay of $D$ blocks.
It is also noteworthy that
the probability of re-transmissions increases
with $M$, i.e., the length of message.

We also have 
similar results in Fig.~\ref{Fig:plt_rtx} (b),
where SNC provides a lower probability of re-transmissions
than $K$-repetition. In particular,
we see that there is no event of more than 2 re-transmissions
with $M \in \{50,100\}$ when SNC is used.
On the other hand, there are cases
that require more than 2 re-transmissions in $K$-repetition.

\begin{figure}[thb]
\begin{center}
\includegraphics[width=\figwidth]{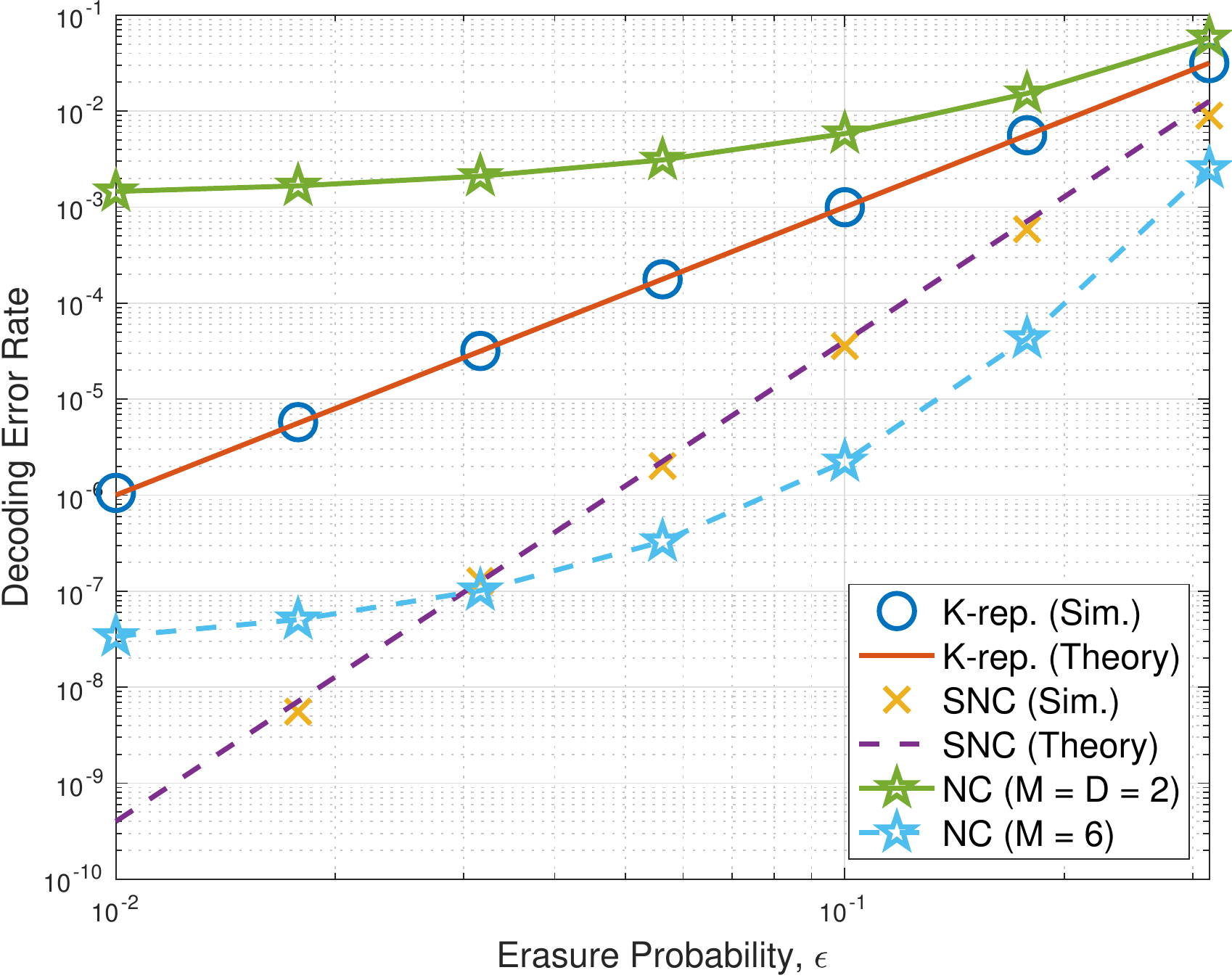} \\
(a) \\
\includegraphics[width=\figwidth]{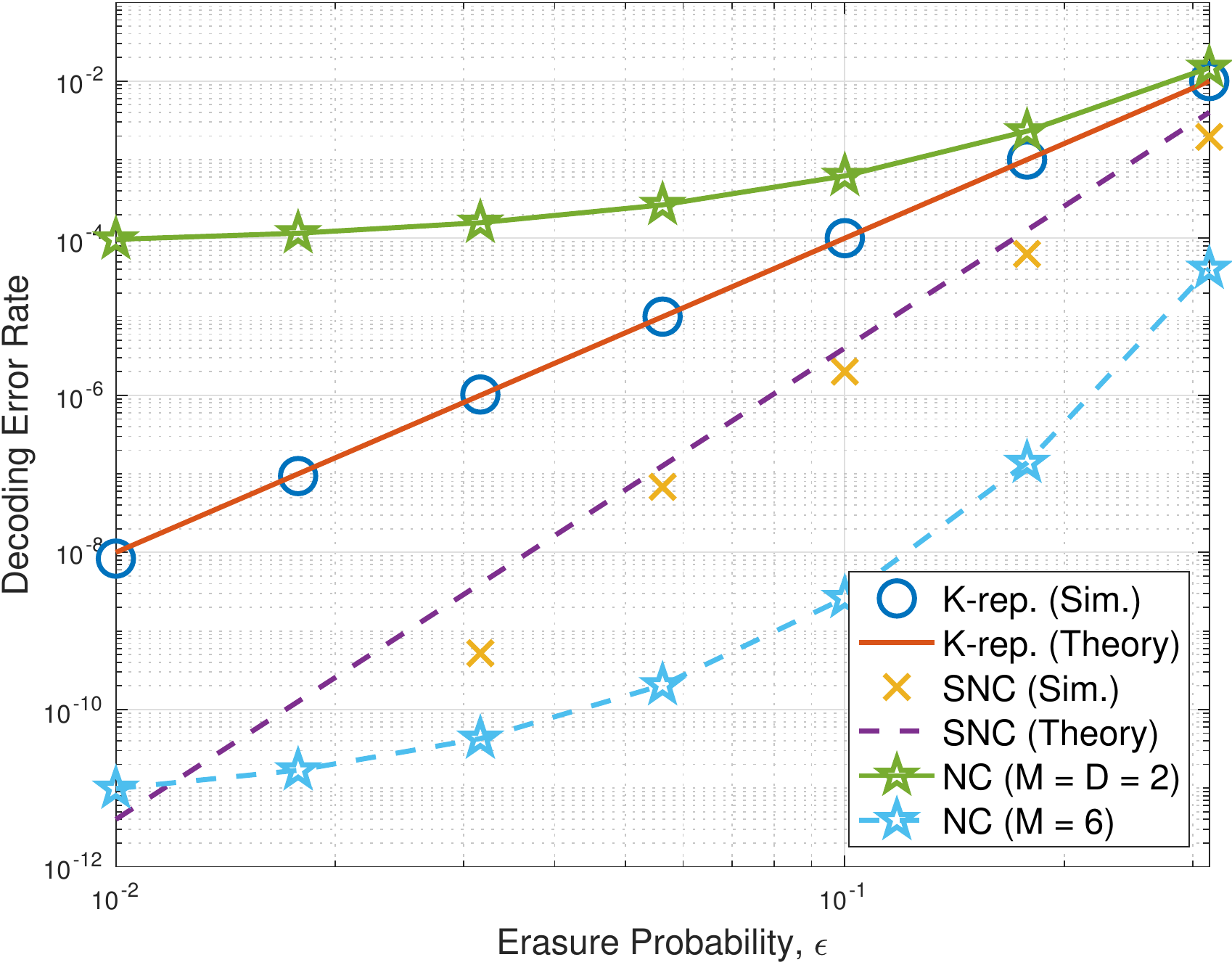} \\
(b) 
\end{center}
\caption{Decoding error rate as a function of erasure probability,
$\epsilon$:
(a) $K = 3$ and $D = 2$;
(b) $K = 4$ and $D = 2$.}
        \label{Fig:plt_eps}
\end{figure}

In order to see the performance for 
different values of erasure probability,
simulations are carried out for $\epsilon 
\in [10^{-2}, 10^{-\frac{1}{2}}]$ and the results
are shown in Fig.~\ref{Fig:plt_eps}.
If $\epsilon$ is too low, we are unable to see any
decoding error events. In this case, the theoretical
prediction from \eqref{EQ:L1} and \eqref{EQ:L3} can be used.

In Fig.~\ref{Fig:plt_eps},
we can see that the decoding error rate increases
with the erasure probability, $\epsilon$.
In addition, as demonstrated earlier,
\eqref{EQ:L1} provides a good prediction
of decoding error rate for $(4,3,2)$-SNC as shown in  
Fig.~\ref{Fig:plt_eps} (a). For $(4,2,2)$-SNC, \eqref{EQ:L3} can be used as an upper-bound,
which can be confirmed by Fig.~\ref{Fig:plt_eps} (b).
By comparing Figs.~\ref{Fig:plt_eps} (a) and (b),
it can be confirmed that the increase of $K$ results in a lower
decoding error rate, and \eqref{EQ:L1} and \eqref{EQ:L3}
are useful to decide $K$ so that a required decoding error rate
can be met
for a given $\epsilon$.

In Fig.~\ref{Fig:plt_eps}, we also include the performance of NC with a fixed $M$. Note that NC is not an on-the-fly scheme, and its total delay (for both encoding and decoding) is $2N = 2MK$ in packets or $2M$ in blocks\footnote{Recall that a block consists of $K$ packets.} 
(each encoding or decoding delay is $M$ in blocks). For a short delay, we can consider the case of $M = D = 2$. From Fig.~\ref{Fig:plt_eps}, we see that NC has a high decoding error rate than both $K$-repetition and SNC. To lower the decoding error rate, a larger $M$, say $M = 6$, can be used for NC. In this case, its performance is comparable to that of SNC. However, the decoding delay of NC becomes $2MK = 36$ (for $K = 3$) or $48$ (for $K = 4$) in packets, while that of SNC is
$K(D+1) = 9$ (for $K = 3$) or $12$ (for $K = 4$) in packets.

\begin{figure}[thb]
\begin{center}
\includegraphics[width=\figwidth]{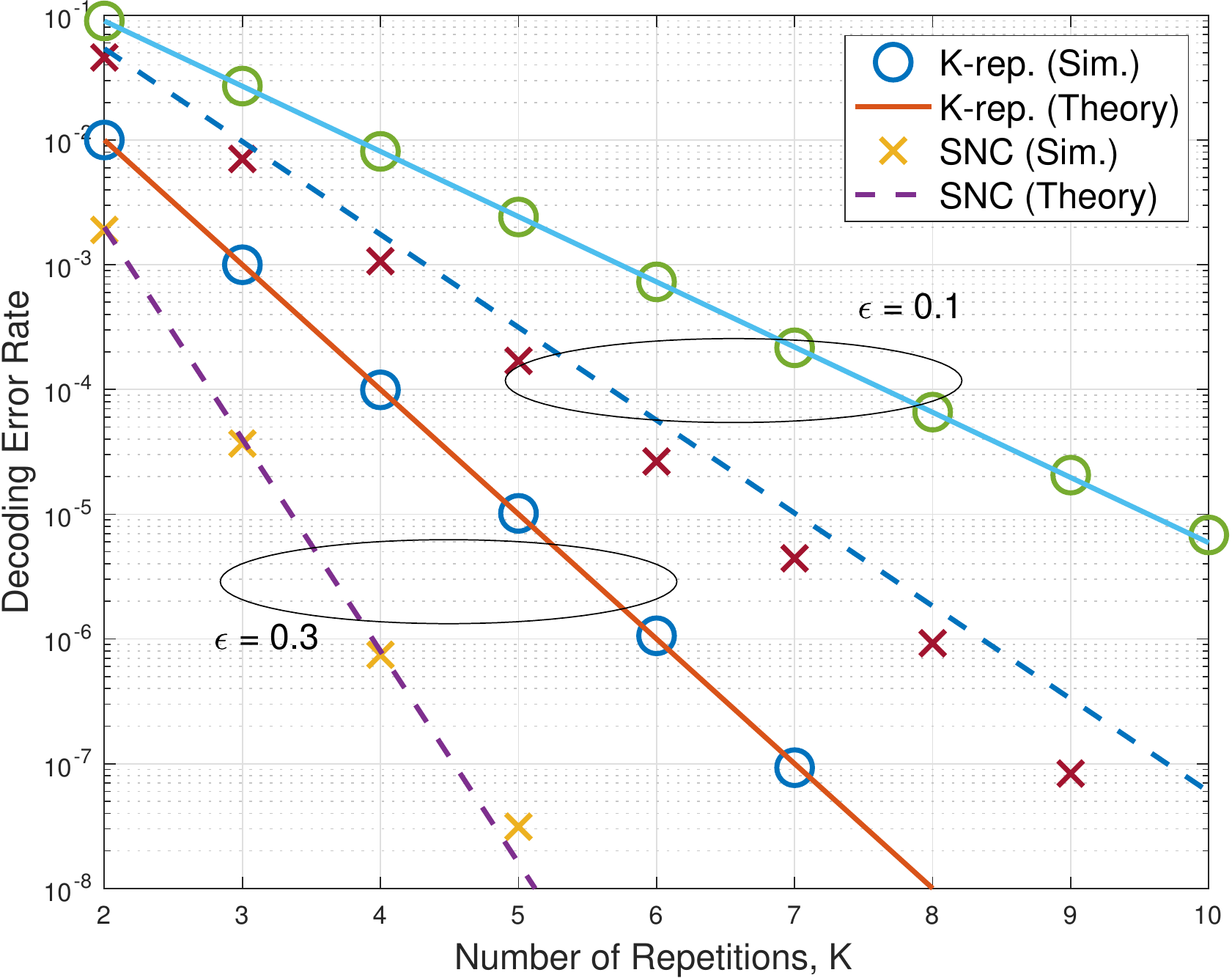} 
\end{center}
\caption{Decoding error rates
of $(K,K-1,2)$-SNC and $K$-repetition as functions of $K$
for a given
erasure probability $\epsilon \in \{0.1, 0.3\}$.}
        \label{Fig:plt_KK1}
\end{figure}

We can have SNC for a different value of $K$ 
as in \eqref{EQ:VKs}, i.e.,
$(K,K-1,2)$-SNC. Fig.~\ref{Fig:plt_KK1}
shows the decoding error rates of
SNC and $K$-repetition
as functions of $K$ for a given
erasure probability $\epsilon \in \{0.1, 0.3\}$.
Clearly, SNC can reduce the number of
repetitions or improve the effective spectral
efficiency, compared to $K$-repetition. For example,
with $\epsilon = 0.1$,
in order to achieve a target decoding error rate
of $10^{-6}$, $K$-repetition
requires $K = 6$ repetitions. On the other hand, 
SNC requires $K = 4$ repetitions. 
Noting that this particular SNC design has a decoding delay
of $K^2 = 16$ slots, while $K$-repetition has a decoding delay
of $K = 6$ slots, we can see that
SNC has about 2.66 times longer decoding delay than 
$K$-repetition,
while the spectral efficiency is improved by a factor of $\frac{6}{4} = 1.5$
in this example.
We can also have similar observations with $\epsilon = 0.3$.
Note that the performance gap in terms of decoding error
probability increases with $K$. Thus, with a lower
target decoding error rate, the performance gap between
SNC and $K$-repetition will be widened.

\section{Concluding Remarks}    \label{S:Con}

In this paper, we proposed SNC to effectively
exploit the performance gain of NC without a significant
increase of decoding delay
for URLLC.
Since
a sliding window of current and past packets 
was used to generate NC packets that are transmitted
together with original packets in SNC as on-the-fly mode,
SNC can be seen as a streaming code. As a result,
SNC is well-suited to the case of URLLC where a transmitter
needs to transmit packets generated at a constant rate
with a high reliability and a guaranteed delay
for each packet delivery.
A few design examples of SNC were also derived
and analyzed. 
It has been shown that the SNC's minimum decoding delay can increase logarithmically with $K$
while its error exponent can be about 2-time larger than
that of $K$-repetition.

While we mainly focused on introducing SNC in this paper with some
design examples,
there are a number of issues to be addressed in the future.
Some of them are as follows.
\begin{itemize}
    \item An optimal design of SNC is necessary. In Lemma~\ref{L:3}, 
    we showed that the error exponent can be greater than
or equal to $2D$. For a given pair of $(K,D)$, there might be an 
optimal design that maximizes the error exponent, which needs to be
investigated in the future.

\item We mainly considered SNC design examples with $q = 2$.
As shown in Lemma~\ref{L:Delay} or \eqref{EQ:DK},
a large $q$ can help decrease the decoding delay. 
Thus, it will be necessary to study SNC with a large $q$.

\end{itemize}

\bibliographystyle{ieeetr}
\bibliography{urllc}
\end{document}